\documentclass[12pt]{article}
\def\tend{\mathop{\to}}

\begin{document}
\begin{center}
\Large \textbf{Nuclear Forces from Chiral Dynamics} \\
\large\textsl{Renat Kh.Gainutdinov and Aigul A.Mutygullina }\\
\textit{Department of Physics,Kazan State University, 18
Kremlevskaya St, Kazan 420008, Russia}
\end{center}
\abstract{

We show that from analysis of time-ordered diagrams for the
two-nucleon $T$-matrix in chiral perturbation theory it follows
that low energy nucleon dynamics is governed by the generalized
dynamical equation, which was derived in [J.Phys.A., v.32, p.5657
(1999)], with a nonlocal-in-time interaction operator. An operator
is presented that parametrizes the nucleon-nucleon interaction in
the ${}^1S_0$ channel at leading order of the Weinberg power
counting.\\

\section{Introduction}

 One of the most fundamental problems of nuclear physics
is to derive nuclear forces from the principles of QCD. A first
attempt to construct a bridge between QCD and low energy nuclear
physics was made by Weinberg \cite{EFT3}. He suggested to derive a
nucleon-nucleon ($NN$) potential in time-ordered chiral
perturbation theory (ChPT). However, such a potential is singular
and the Schr{\"o}dinger (Lippmann-Schwinger) equation makes no
sense without regularization and renormalization. Thus in the
effective field theory (EFT) of nuclear forces, which following
the pioneering work of Weinberg has become very popular in nuclear
physics (for a review, see Ref.\cite{rev}), the Schr{\"o}dinger
equation is not valid. On the other hand, quantum mechanics is one
of the basic ingredients of the whole formalism of fields and
particles, and hence in the nonrelativistic limit QCD must produce
low energy nucleon physics consistent with the basic principles of
quantum mechanics. However, as it follows from the Weinberg
analysis, QCD leads, through ChPT, to the low energy theory in
which the Schr{\"o}dinger equation is not valid. This means that
either there is something wrong with QCD and ChPT or the
Schr{\"o}dinger equation is not the basic dynamical equation of
quantum theory. Meanwhile, in Ref.\cite{R.Kh.:1999} it has been
shown that the Schr{\"o}dinger equation is not the most general
equation consistent with the current concepts of quantum physics,
and a more general equation of motion has been derived as a
consequence of the basic postulates of the Feynman
\cite{Feynman:1948} and canonical approaches to quantum theory.
Being equivalent to the Schr{\"o}dinger equation in the case of
instantaneous interactions, this generalized dynamical equation
permits the generalization to the case where the dynamics of a
system is generated by a nonlocal-in-time interaction. The
generalized quantum dynamics (GQD) developed in this way has
proved an useful tool for solving various problems in quantum
theory \cite{PRC:2002,PLA:2002}.

The Weinberg program for low energy nucleon physics employs the
analysis of time-ordered diagrams for the two-nucleon $T$-matrix
in ChPT to derive a $NN$ potential and then to use it in the
Lippmann-Schwinger (LS) equation for constructing the full $NN$
$T$-matrix. Obviously the starting point for this program is the
assumption that in the nonrelativistic limit ChPT leads to low
energy nucleon dynamics which is Hamiltonian and is governed by
the Schr{\"o}dinger equation. However, the fact that the chiral
potentials constructed in this way are singular and result in UV
divergences mean that this assumption has not corroborated.

The GQD provides a new insight into this problem: The above may
mean that the low energy nucleon dynamics, which results from the
analysis of diagrams in ChPT, is governed by the generalized
dynamical equation with a nonlocal-in-time interaction operator
when this equation cannot be reduced to the Schr{\"o}dinger
equation. In Ref.\cite{PRC:2002} it has been shown that such a
dynamical situation takes place in the case of the pionless EFT:
After renormalization in leading order low energy nucleon dynamics
is governed by nonlocal-in-time interaction operator. Moreover,
this dynamics is just the same as in the model
\cite{R.Kh.:1999,R.Kh./A.A.:1999} developed as a test model
illustrating the possibility of the extension of quantum dynamics
provided by the formalism of the GQD. This gives us the expect
that, if we take into account that low energy nucleon dynamics
need not be governed by the Schr{\"o}dinger equation, we will be
able to derive from the analysis of diagrams for the two-nucleon
$T$-matrix in ChPT a well defined $NN$ interaction operator which
will not give rise to UV divergences. In the present paper we will
show that from the analysis of these diagrams it really follows
that the effective $NN$ interaction generating low energy nucleon
dynamics is nonlocal in time. In leading order of the Weinberg
power counting we will construct a chiral $NN$ interaction
operator for the ${}^1S_0$ channel. It will be shown that the
generalized dynamical equation with this interaction operator is
well defined and allows one to construct the $T$-matrix and the
evolution operator without regularization and renormalization.

\section{Generalized Quantum Dynamics}

In the GQD the evolution operator $U(t,t_0)$ in the interaction
picture is represented in the form \cite{R.Kh.:1999}
\begin{equation}
<\psi_2| U(t,t_0)|\psi_1> = <\psi_2|\psi_1> + \int_{t_0}^t dt_2
\int_{t_0}^{t_2} dt_1 <\psi_2|\tilde S(t_2,t_1)|\psi_1>,
\label{repre}
\end{equation}
where $<\psi_2|\tilde S(t_2,t_1)|\psi_1>$ is the probability
amplitude that if at time $t_1$ the system was in the state
$|\psi_1>,$ then the interaction in the system will begin at time
$t_1$ and will end at  time $t_2,$ and at this time the system
will be in the state $|\psi_2>.$ This equation represents the
Feynman superposition principle  according to which the
probability amplitude of an event which can happen in several
different ways is a sum of contributions from each alternative
way. Here subprocesses with definite instants of the beginning and
end of the interaction in the system are used as alternative ways
of realization of the corresponding evolution process, and
$<\psi_2|\tilde S(t_2,t_1)|\psi_1>$ represents the contribution to
the evolution operator from the subprocess in which the
interaction begins at time $t_1$ and ends at  time $t_2$.  As has
been shown in Ref. \cite{R.Kh.:1999}, for the evolution operator
given by Eq.(\ref{repre}) to be unitary for any $t$ and $t_0$ the
operator $\tilde S(t_2,t_1)$ must satisfy the equation
\begin{eqnarray}
(t_2-t_1) \tilde S(t_2,t_1) 
= \int^{t_2}_{t_1} dt_4 \int^{t_4}_{t_1}dt_3 (t_4-t_3)\tilde
S(t_2,t_4) \tilde S(t_3,t_1). \label{main}
\end{eqnarray}
A remarkable feature of this relation  is that it works as a
recurrence relation and allows one to obtain the operators $\tilde
S(t_2,t_1)$ for any $t_1$ and $t_2$, if $\tilde S(t'_2, t'_1)$
corresponding to infinitesimal duration times $\tau = t'_2 -t'_1$
of interaction are known. It is natural to assume that most of the
contribution to the evolution operator in the limit $t_2\to t_1$
comes from the processes associated with the fundamental
interaction in the system under study. Denoting this contribution
by $H_{int}(t_2,t_1)$ we can write
\begin{equation}
\tilde{S}(t_2,t_1) \tend\limits_{t_2\rightarrow t_1}
H_{int}(t_2,t_1) + o(\tau^{\epsilon}),\label{gran}
\end{equation}
where $\tau=t_2-t_1$. The parameter $\varepsilon$ is determined by
demanding that $H_{int}(t_2,t_1)$  called the generalized
interaction operator must be so close to the solution of
Eq.(\ref{main}) in the limit $t_2\tend t_1$ that this equation has
a unique solution having the behavior (\ref{gran}) near the point
$t_2=t_1$. If $H_{int}(t_2,t_1)$ is specified, Eq.(\ref{main})
allows one to find the operator $\tilde S(t_2,t_1)$, and hence the
evolution operator. Thus Eq.(\ref{main}) which is a direct
consequence of the principle of the superposition can be regarded
as an equation of motion for states of a quantum system.  This
equation allows one to construct the evolution operator by using
the contributions from fundamental processes as building blocks.
In the case of Hamiltonian dynamics the fundamental interaction is
instantaneous. The generalized interaction operator describing
such an interaction is of the form
\begin{equation}
 H_{int}(t_2,t_1) = - 2i \delta(t_2-t_1)
 H_{I}(t_1) \label{loc}
\end{equation}
(the delta function $\delta(t_2-t_1)$ emphasizes that the
interaction is instantaneous). In this case Eq.(\ref{main}) is
equivalent to the Schr{\"o}dinger equation \cite{R.Kh.:1999}, and
the operator $H_I(t)$ is an interaction Hamiltonian. At the same
time, Eq.(\ref{main}) permits the generalization to the case where
the fundamental interaction in
 a quantum system is nonlocal in time, and hence
the dynamics is non-Hamiltonian \cite{R.Kh.:1999}.

 By
using Eq.(\ref{repre}), for $U(t,t_0)$, we can write
\begin{eqnarray}
U(t,t_0)=  {\bf 1} + \frac{i}{2\pi}
\int^\infty_{-\infty} dx\exp[-i(z-H_0)t]\nonumber\\
\times (z-H_0)^{-1}T(z)(z-H_0)^{-1} \exp[i(z-H_0)t_0], \label{evo}
\end{eqnarray}
 where $z=x+iy$, $y>0$, and $H_0$ is the free Hamiltonian.
 The
 operator $T(z)$ is defined by
\begin{eqnarray}
T(z) = i \int_{0}^{\infty} d\tau \exp(iz\tau)\tilde
T(\tau),\label{tz}
\end{eqnarray}
where $\tilde T(\tau)=\exp(-iH_0t_2)\tilde
S(t_2,t_1)\exp(iH_0t_1)$, and $\tau=t_2-t_1$.
 In terms of the
T-matrix defined by Eq.(\ref{tz}) the generalized dynamical
equation (\ref{main}) can be rewritten in the form
\cite{R.Kh.:1999}
\begin{equation}
\frac{d \langle \psi_2|T(z)|\psi_1\rangle}{dz} = -  \sum
\limits_{n} \frac{\langle \psi_2|T(z)|n\rangle\langle
n|T(z)|\psi_1\rangle}{(z-E_n)^2},\label{difer}
\end{equation}
where $n$ stands for the entire set of discrete and continuous
variables that characterize the system in full, and $|n\rangle$
are the eigenvectors of $H_0$. As it follows from Eq. (\ref{loc}),
the boundary condition for this equation is of the form
\begin{equation}
\langle\psi_2|T(z)|\psi_1\rangle \tend
\limits_{|z|\tend\infty}\langle\psi_2|B(z)|\psi_1\rangle+o(|z|^{-\beta}),\label{as}
\end{equation}
where
\begin{equation}
B(z)= i \int_{0}^{\infty} d\tau \exp(iz\tau) H_{int}^{(s)}(\tau),
\end{equation}
and $H^{(s)}_{int}(t_2-t_1) = \exp(-iH_0t_2) H_{int}(t_2,t_1)
\exp(iH_0t_1)$ is the interaction operator in the Schr{\"o}dinger
picture. As can be seen from Eqs.(8) and (9), the operator $B(z)$
represents the contribution which $H^{(s)}_{int}(\tau)$ gives to
the operator $T(z)$.

\section{Chiral Dynamics and the $NN$ Interaction Operator}

The Weinberg program implies to employ the analysis of diagrams in
ChPT for deriving the chiral $NN$ potential which is assumed to be
a sum of irreducible diagrams involving only two external
nucleons. Here irreducible diagrams are two-nucleon irreducible:
Any intermediate state contains at least one pion or isobar. The
motivation for this is as follows. If we assume that the
two-nucleon $T$-matrix satisfies the $LS$ equation with the
interaction potential $\langle{\bf p}_2|V|{\bf p}_1\rangle$, then
in the limit $|z|\to\infty$ the operator $T(z)$ must behave as
\begin{equation}
\langle{\bf p}_2|T(z)|{\bf
p}_1\rangle\tend\limits_{|z|\tend\infty}\langle{\bf p}_2|V|{\bf
p}_1\rangle .
\end{equation}
Since any reducible diagram can be constructed from irreducible
ones by connecting the latter with intermediate two-nucleon states
whose evolution is described by the free Green operator $
G_0(z)=(z-H_0)^{-1}$, reducible diagrams tend to zero as
$|z|\to\infty$ and hence do not contribute to the potential
$\langle{\bf p}_2|V|{\bf p}_1\rangle$. For this reason it is
natural to use the sum of irreducible diagrams as a potential
describing a "fundamental" interaction in the two-nucleon system.
In addition, as it follows from Eq.(\ref{tz}), the large $z$
behavior of the $T$-matrix relates to the behavior of the operator
$\tilde S(t_2,t_1)$ in the limit of infinitesimal duration times
of interaction in the system when this operator describes a
fundamental interaction in the system. In leading order of
Weinberg power counting we have two types of irreducible diagrams:
The one-pion-exchange diagram and the diagram describing the
contact interaction. The corresponding chiral potential is of the
form
\begin{equation}
\langle{\bf p}_2|V|{\bf p}_1\rangle=V_{OPE}({\bf p}_2,{\bf
p}_1)+C_S+C_T{\bf{\sigma}_1}\cdot{\bf{\sigma}_2},\label{ope}
\end{equation}
where $V_{OPE}({\bf p}_2,{\bf p}_1)$ is the one-pion exchange
(OPE) potential
$$V_{OPE}({\bf p}_2,{\bf
p}_1)=-\left(\frac{g_A^2}{2f_\pi^2}\right)\frac{\bf{q}
\cdot\sigma_1\bf{q} \cdot{\bf \sigma}_2}
{q^2+m_\pi^2}\bf{\tau}_1\cdot\bf{\tau}_2,$$ with ${\bf
q}\equiv{\bf p}_2-{\bf p}_1$. The coupling $g_A$ is the axial
coupling constant, $m_\pi$ is the pion mass, $f_\pi$ is the pion
decay constant, and $\sigma(\tau)$ are the Pauli matrices acting
in spin (isospin) space. In the ${}^1S_0$ channel the chiral
potential can be rewritten in the form
\begin{equation}
\langle{\bf p}_2|V|{\bf p}_1\rangle=C_0+V_{\pi}({\bf p}_2,{\bf
p}_1),
\end{equation}
where $C_0=C_S-3C_T$ and
\begin{equation}
V_{\pi}({\bf p}_2,{\bf p}_1)=-\frac{4\pi\alpha_\pi}
{q^2+m_\pi^2},\qquad\alpha_\pi=\frac{g_A^2m_\pi^2}{8\pi f_\pi^2}.
\end{equation}
However, the potential (\ref{ope}) is singular and the
Schr{\"o}dinger equation makes no sense without regularization and
renormalization.The reason for this is that the chiral Lagrangian
is only of formal importance. In addition, one has to specify a
renormalization scheme to make the physical predictions finite.
 The formal chiral Lagrangian determines the structure of
the theory. In particular it determines the structure of the
two-nucleon $T$-matrix, i.e., its dependence on the momenta of
nucleons. On the other hand, this $T$-matrix must satisfy
Eq.(\ref{difer}), and there is the one-to-one correspondence
between the large-momentum behavior of the $T$-matrix and the
character of the dynamics in the two-nucleon system
\cite{R.Kh.:1999}. As will be shown below, in order to establish
such a behavior one need not to enter into detail and perform
renormalization of diagrams: This behavior directly follows from
the form of the chiral Lagrangian. This gives us the hope that,
using the structure of the two-nucleon diagrams which follows from
ChPT together with the requirement that the $T$-matrix satisfy the
generalized dynamical equation may allow one to find a true
asymptotic behavior for $T(z)$. By using this behavior, one can
then derive the $NN$ interaction operator. With such an operator
the generalized dynamical equation permits one to construct the
two-nucleon $T$-matrix without resorting to regularization and
renormalization.

The form of the chiral Lagrangian specifies the dependence of the
two-nucleon $T$-matrix on the momenta of nucleons. From the
analysis of the time-ordered diagrams in ChPT it follows that the
leading order two-nucleon $T$-matrix should be of the form
\begin{equation}
\langle{\bf p}_2|T(z)|{\bf p}_1\rangle=t_{00}(z)+t_{01}(z,{\bf
p}_1)+ t_{01}(z,{\bf p}_2)+ t_{11}(z,{\bf p}_1,{\bf
p}_2).\label{sum}
\end{equation}
In fact, there are two kinds of vertices in the leading order
diagrams: The vertex which corresponds to the contact interaction
without derivatives and is a constant, and the $\pi NN$ vertex.
The term $t_{00}(z)$ describes the contribution from the
time-ordered diagrams in which the initial and final vertices are
the above contact ones. The term $ t_{01}(z,{\bf p}_1)$ is the
contribution from the diagrams in which the initial and final
vertices are the $\pi NN$ and contact ones respectively.
Correspondingly the initial and final vertices in the diagrams
which contribute to the term $ t_{11}(z,{\bf p}_1,{\bf p}_2)$ are
the $\pi NN$ ones. Thus Eq. (\ref{sum}) represents the structure
of the leading order two-nucleon $T$-matrix which results from the
form of the chiral Lagrangian. As has been noted, the character of
the dynamics of a system is determined by the behavior of $T(z)$
in the limit $|z|\to\infty$. Let us examine this behavior of the
two-nucleon $T$-matrix (\ref{sum}) using the fact that this
$T$-matrix must satisfy Eq.(\ref{difer}).  The main contribution
to $t_{11}(z,{\bf p}_1,{\bf p}_2)$ in the limit $|z|\to\infty$
comes from the two-nucleon irreducible diagram describing the
one-pion-exchange processes, and, for the ${}^1S_0$ channel, we
have
\begin{equation}
t_{11}(z,{\bf p}_1,{\bf p}_2)\tend\limits_{|z|\to\infty}V_\pi
({\bf p}_1,{\bf p}_2).\label{ob}
\end{equation}
Since there no irreducible diagrams which contribute to the terms
$t_{01}(z,{\bf p})$ and $ t_{10}(z,{\bf p}_1)$, it is natural to
assume that  $\lim\limits_{|z|\to\infty} t_{01}(z,{\bf
p})=\lim\limits_{|z|\to\infty} t_{10}(z,{\bf p})=0.$ At the same
time, one might think that in the limit $|z|\to\infty$ the
function $t_{00}(z)$ tends to a nonzero constant $C_0$ playing the
role of a contact potential. However, such a potential is singular
and leads to UV divergences. This is a manifestation of the fact
that the large-momentum behavior of the $T$-matrix of the form
(\ref{sum}) does not satisfy the requirements of Hamiltonian
formalism. There are no potentials which could generate the
$T$-matrix of such a form, and hence the dynamics of the system is
non-Hamiltonian. On the other hand, the form (\ref{sum}) of the
two-nucleon $T$-matrix is not at variance with the requirements of
the GQD. Such a form of the $T$-matrix corresponds to the case of
a nonlocal-in-time interaction when the dynamics is
non-Hamiltonian. In this case the contact term must tend to zero
as $|z|\to\infty$ \cite{PRC:2002}.

The large $z$ behavior of $T(z)$  is determined by
Eq.(\ref{difer}). From this equation it follows that $t_{00}(z)$
and $t_{ij}(z,{\bf p})$ satisfy the equations
\begin{eqnarray}
\frac {dt_{00}(z)}{dz} = - \int \frac{d^3k}{(2\pi)^3} \frac
{\left(t_{00}(z)+t_{01}(z,{\bf
k})\right)\left(t_{00}(z)+t_{10}(z,{\bf k})\right)}
 {(z-E_k)^2},
 \label{deq00}
\end{eqnarray}
\begin{eqnarray}
\frac {dt_{01}(z,{\bf p})}{dz} =- \int \frac{d^3k}{(2\pi)^3} \frac
{\left(t_{00}(z)+t_{01}(z,{\bf k})\right)\left(t_{01}(z,{\bf
p})+t_{11}(z,{\bf k},{\bf p} )\right)} {(z-E_k)^2},
 \label{deq01}
\end{eqnarray}
\begin{eqnarray}
\frac {dt_{10}(z,{\bf p})}{dz} = - \int \frac{d^3k}{(2\pi)^3}
\frac {\left(t_{10}(z,{\bf p})+t_{11}(z,{\bf p},{\bf k}
)\right)\left(t_{00}(z)+t_{10}(z,{\bf k})\right)} {(z-E_k)^2}.
\label{deq10}
\end{eqnarray}
Since $t_{01}(z,{\bf p})$ and $t_{10}(z,{\bf p})$ describe the
contribution form diagrams which consist of connected diagrams
corresponding to the terms $t_{00}(z)$ and $t_{11}(z,{\bf
p}_1,{\bf p}_2)$, in the limit $|z|\to\infty$ the terms
$t_{01}(z,{\bf p})$ and $t_{10}(z,{\bf p})$ must tend to zero
faster than the term $t_{00}(z)$. Taking  this fact into account,
from Eq.(\ref{deq00}) we get
$$
\frac {dt_{00}(z)}{dz} \sim -(t_{00}(z))^2 \int
\frac{d^3k}{(2\pi)^3} \frac {1}{(z-E_k)^2},\qquad |z|\to\infty. $$
From this equation, in turn, it follows that
\begin{equation}
t_{00}(z)=b_1(-z)^{-1/2}+o(|z|^{-1/2}),\qquad |z|\to\infty,
\end{equation}
where $b_1=-\frac{4\pi}{M\sqrt{M}}$. Substituting this expression
into Eq.(17) and taking into account Eq.(\ref{ob}) yield
\begin{eqnarray}
\frac {dt_{01}(z,{\bf p})}{dz} = - \frac{b_1}{\sqrt{-z}}\int
\frac{d^3k}{(2\pi)^3} \frac
{(V_\pi({\bf k},{\bf p})+t_{01}(z,{\bf p}))} {(z-E_k)^2}
+o(|z|^{-2}). \nonumber
\end{eqnarray}
From this equation we get
$$
t_{01}(z,{\bf p}) =B_{01}(z,{\bf p}) +o(|z|^{-1}),\qquad
|z|\to\infty, $$
 where $$B_{01}(z,{\bf p})=
\frac{b_1}{\sqrt{-z}}\int \frac{d^3k}{(2\pi)^3} \frac {V_\pi({\bf
k},{\bf p})} {(z-E_k)}.$$ In the same way, for $t_{10}(z,{\bf
p})$, we get $ t_{10}(z,{\bf p}) =B_{10}(z,{\bf p}) +o(|z|^{-1}),
$ where $B_{10}(z,{\bf p})=B_{01}(z,{\bf p})$. Equation (19)
represents the first term in the asymptotic expansion of
$t_{00}(z)$. In order to obtain next terms in this expansion, let
us write the equation for $\tilde t_{00}(z)\equiv
t_{00}(z)-b_1(-z)^{-1/2}$
\begin{eqnarray}
\frac {d\tilde t_{00}(z)}{dz} = -\frac{2b_1^
2}{z}\int\frac{d^3k_1d^3k_2}{(2\pi)^6} \frac {V_\pi({\bf k}_1,{\bf
k}_2)}
{(z-E_{k_1})^2(z-E_{k_2})}\nonumber\\
-\frac{2b_1}{\sqrt{-z}}\int \frac{d^3k}{(2\pi)^3} \frac {\tilde
t_{00}(z)}{(z-E_k)^2}+o(|z|^{-2})
=-\frac{\tilde t_{00}(z)}{z}- \frac{b_3}{z^2}
+o(|z|^{-2}),
\end{eqnarray}
where $$b_3=-2b_1^ 2
4\pi\alpha_\pi\int\frac{d^3q_1d^3q_2}{(2\pi)^6} ({\bf q}_1-{\bf
q}_2)^{-2}(1-q_1^2/M)^{-2}(1-q_2^2/M)^{-1},$$ with ${\bf q}_i={\bf
p}_i/\sqrt{z}$. Here we have used Eq.(\ref{deq00}) and the above
expressions for $t_{00}(z)$, $t_{01}(z,{\bf p})$ and
$t_{10}(z,{\bf p})$.
 The solution of Eq.(20) is of the form
$ \tilde t_{00}(z)=-\frac{b_2}{z}-b_3\frac{\ln z}{z}+o(|z|^{-1}),
$ where $b_2$ is a free parameter, and hence
$$
t_{00}(z)=\frac{b_1}{\sqrt{-z}}-\frac{b_2}{z}-b_3\frac{\ln
z}{z}+o(|z|^{-1}),\qquad |z|\to\infty.$$
 Thus the
requirement that the $T$-matrix be of the form (\ref{sum}) and
satisfy the generalized dynamical equation determines its behavior
for $|z|\to\infty$ up to one free parameter $b_2$
\begin{eqnarray}
\langle{\bf p}_2|T(z)|{\bf
p}_1\rangle=\frac{b_1}{\sqrt{-z}}-\frac{b_2}{z}-b_3\frac{\ln
z}{z}+ B_{01}(z,{\bf p}_1)+B_{10}(z,{\bf
p}_2)\nonumber\\
+V_\pi({\bf p}_1,{\bf p}_2)+o(|z|^{-1}), \qquad
|z|\to\infty.
\end{eqnarray}
Knowing the terms in the asymptotic expansion of $\langle{\bf
p}_2|T(z)|{\bf p}_1\rangle$ represented in Eq.(21) is sufficient
to construct the $T$-matrix, and hence they determine the
effective interaction operator, which governs the dynamics of a
two-nucleon system. Thus in the ${}^1S_0$ channel this operator is
of the form
\begin{equation} \langle{\bf p}_2|B(z)|{\bf
p}_1\rangle=\frac{b_1}{\sqrt{-z}}-\frac{b_2}{z}-b_3\frac{\ln
z}{z}+B_{01}(z,{\bf p}_1)+B_{10}(z,{\bf p}_2) +V_\pi({\bf
p}_1,{\bf p}_2).\label{os}
\end{equation}
The corresponding $NN$ interaction operator $H_{int}^{(s)}(\tau)$
can be obtained by using Eq.(9). The generalized dynamical
equation with this nonlocal-in-time interaction operator is well
defined and allows one to construct the $T$-matrix without
regularization and renormalization. It can be shown that the
solution of Eq.(\ref{difer}) with this interaction operator is of
the form
\begin{equation}
\langle{\bf p}_2|T(z)|{\bf
p}_1\rangle=t_L(z,{\bf p}_1,{\bf p}_2)+\chi(z,{\bf
p}_2)\chi(z,{\bf p}_1)\left[C^{-1}-G_z^{(R)}(0,0)\right]^{-1},
\end{equation}
where
\begin{eqnarray}
G_z^{(R)}(0,0)=(4\pi)^{-1} M^{
\frac{3}{2}}{\sqrt{-z}}-2\int_0^zds\int\frac{d^3k_1d^3k_2}{(2\pi)^6}
\frac {V_\pi({\bf k}_1,{\bf k}_2)}
{(s-E_{k_1})^2(s-E_{k_2})}\nonumber\\
+\int\frac{d^3k_1d^3k_2d^3k_3}{(2\pi)^9} \frac {V_\pi({\bf
k}_1,{\bf k}_2)T_L(z,{\bf k}_2,{\bf k}_3)}
{(z-E_{k_1})(z-E_{k_2})(z-E_{k_3})},\nonumber\\
\chi(z,{\bf p})=1+\int\frac{d^3k}{(2\pi)^3} \frac {T_L(z,{\bf
p},{\bf k})} {(z-E_{k})},\nonumber
\end{eqnarray}
$C=-(4\pi)^2M^{-3}b_2^{-1},$and $T_L(z,{\bf k}_2,{\bf k}_1)$ is
the solution of the $LS$ equation with the Yukawa potential
$V_\pi({\bf k}_2,{\bf k}_1)$. For the scattering amplitude, we
then get
\begin{equation}
A(p)=A_\pi-\left\{\chi(E_p,{\bf
p})\right\}^2\left[C^{-1}-G^{(R)}_{E_p}(0,0)\right]^{-1},\label{A}
\end{equation}
where $A_\pi$ is just the amplitudes one finds in the pure Yukawa
theory with the potential $V_\pi({\bf p}_2,{\bf p}_1)$. This
expression for the scattering amplitude is just the same that has
been obtained by Kaplan, Savage and Wise \cite{Kaplan2} by summing
and renormalizing the relevant diagrams in the EFT of nuclear
forces. Equations (\ref{evo}) and (23), can then be used for
constructing the evolution operator describing the dynamics of the
two-nucleon system in the ${}^1S_0$ channel.

\section{Conclusion}

We have shown that the GQD provides a new way to derive the low
energy theory of the $NN$ interaction from ChPT. In this way one
can use the structure of the theory predicted by ChPT together
with the requirement that the resultant two-nucleon $T$-matrix
satisfy the generalized dynamical equation. For example, as it
follows from the analysis of diagrams in ChPT, in the ${}^1S_0$
channel the leading order two-nucleon $T$-matrix should be of the
form (\ref{sum}).  As we have shown, such a form of the $T$-matrix
corresponds to the case where the dynamics in a quantum system is
governed by a nonlocal-in-time interaction. Thus from the analysis
of time-ordered diagrams for the two-nucleon $T$-matrix in ChPT it
follows that low energy nucleon dynamics is governed by the
generalized dynamical equation with a nonlocal-in-time operator of
the $NN$ interaction. We have shown that the requirement that
two-nucleon $T$-matrix (\ref{sum}) satisfy Eq.(\ref{difer}) and
the boundary condition (\ref{ob}) determines this operator up to
one arbitrary parameter. In the ${}^1S_0$ channel the leading
order $NN$ interaction operator is of the form (\ref{os}). The
generalized dynamical equation with this interaction operator is
well defined and allows one to construct the $T$-matrix and the
evolution operator without resorting to regularization and
renormalization procedures.

Finally, we have shown that the formalism of the GQD provides a
new way to realize the Weinberg program for deriving the forces
between nucleons from the analysis of diagrams in ChPT. It is
hopped that in this way one can construct a perfectly consistent
theory of nuclear forces free from UV divergences.

\end{document}